\documentclass[aps,pra,twocolumn,amsmath,amssymb,nofootinbib,showpacs,superscriptaddress]{revtex4-1}
\usepackage[english]{babel}
\usepackage{latexsym}
\usepackage{graphics}
\usepackage{graphicx}
\usepackage{epsfig}
\usepackage{color}
\usepackage{bm}
\usepackage{amsmath}
\usepackage{amssymb}
\usepackage{amsthm}
\usepackage{dcolumn}
\usepackage{bm}
\usepackage{float}
\usepackage{hyperref}
\usepackage{color}
\usepackage{epstopdf}
\usepackage{cleveref}
\usepackage[svgnames]{xcolor}
\usepackage{enumerate}
\usepackage{braket}
\hypersetup{hidelinks,colorlinks=true,allcolors=DarkBlue}

\theoremstyle{remark}

\begin{document}

\preprint{APS/123-QED}

\title{Learning phase transitions in ferrimagnetic GdFeCo alloys}

\author{N.A. Koritsky}
\affiliation{Russian Quantum Center, Skolkovo, Moscow 143025, Russia}
\affiliation{Moscow Institute of Physics and Technology, Dolgoprudny 141700, Russia}

\author{S.V. Solov'yov}
\affiliation{Moscow Institute of Physics and Technology, Dolgoprudny 141700, Russia}

\author{A.K. Fedorov}
\affiliation{Russian Quantum Center, Skolkovo, Moscow 143025, Russia}
\affiliation{Moscow Institute of Physics and Technology, Dolgoprudny 141700, Russia}

\author{A.K. Zvezdin}
\affiliation{Russian Quantum Center, Skolkovo, Moscow 143025, Russia}
\affiliation{Moscow Institute of Physics and Technology, Dolgoprudny 141700, Russia}
\affiliation{Prokhorov General Physics Institute of the Russian Academy of Sciences, Moscow  119991, Russia}

\date{\today}
\begin{abstract}
We present results on the identification of phase transitions in ferrimagnetic GdFeCo alloys using machine learning. 
The approach for finding phase transitions in the system is based on the `learning by confusion' scheme, which allows one to characterize phase transitions using a universal $W$-shape. 
By applying the `learning by confusion' scheme, we obtain 2D $W$-a shaped surface that characterizes a triple phase transition point of the GdFeCo alloy. 
We demonstrate that our results are in the perfect agreement with the procedure of the numerical minimization of the thermodynamical potential, 
yet our machine-learning-based scheme has the potential to provide a speedup in the task of the phase transition identification. 
\end{abstract}

\maketitle

\section{Introduction}

The study of phases in condensed-matter systems is of paramount importance for elucidating the underlying physical principles behind their properties.
At the same time, condensed-matter systems are considered as platforms for realizing various technologies. 
Among a rich family of magnetic materials, rare-earth amorphous alloys and intermetallics attract special attention.
They are considered as a promising systems for various fields, 
such as spintronics~\cite{DasSarma2004}, magnonics~\cite{Demokritov2010}, and ultrafast magnetism~\cite{Rasing2007,Rasing2009,Fullerton2002,Ostler2012,Rasing2012,Radu2011,Graves2013,Stanciu2007}.
However, the complexity of such systems makes the task of the analysis of their phases very challenging. 
Except for very few cases, analytical methods cannot be used for understanding their properties.
Numerical methods are not universally helpful since they typically encounter technical difficulties. 

Recently, machine learning techniques have been considered as a potential tool for analyzing complex systems both in classical and quantum domains~\cite{Carleo2019}.
One of the approaches is to encode quantum states in the parameters of neural networks, such as restricted Boltzmann machines.
This idea has been explored in the analysis of many-body quantum models, such as spin-$1/2$ systems~\cite{Troyer2017}, and for quantum tomography~\cite{Troyer2018,Tiunov2020}. 
An alternative approach is to use machine learning for the analysis of state configurations samples. 
This methodology has been successfully used for discriminating phase transitions in the correlated many-body system.
An elegant trick that can be incorporated in this approach is to use `learning by confusion' scheme~\cite{Nieuwenburg2017} for revealing phase transitions. 
By having a-priori knowledge that the system has two distinct phases, where the transition is controlled by a parameter, we train a neural network by proposed some values of the critical parameter and evaluate network's performance. 
It is likely to expect that the best performance of the network takes place for the case, where the critical parameter is chosen correctly. 
In this case, the transition is characterized by universal W-shape.
The `learning by confusion'  scheme has been studied for many-body localization--delocalization transition~\cite{Gornyi2018}, 2D percolation and Ising models~\cite{Zhao2019}, 
critical behavior of the two-color Ashkin-Teller model, the XY model, and the eight-state clock model~\cite{Lee2019}, exploring topological states~\cite{Granath2019}, and finding chaos-integrability transition~\cite{Kharkov2020}. 
The potential of machine learning can be then used to explore interesting properties of magnetic systems of various kinds.

The ferrimagnetic GdFeCo alloy is a particular example of such an amorphous alloy, which is a promising material both from the side of fundamental research and potential applications.
It has been shown that there is a possibility of ultra-fast switching magnetization by a femtosecond laser~\cite{Guyader2012, Graves2013}, which is promising for data storage.
In addition, the widely varying stoichiometric composition provides opportunities to control the magnetic and spin transport properties of GdFeCo films~\cite{Kawakami2020, Roschewsky2017}.
In the presence of a perpendicular oriented external magnetic field, the GdFeCo alloy demonstrates the presence 
of a triple phase transition point with respect to the change of the external magnetic field ($H$) and the stoichiometric coefficient in the Gd\textsubscript{x}(FeCo)\textsubscript{1-x} ($x$).
In the vicinity of such points magnetization, metastable states may be observed~\cite{Davydova2019}, which can be detected by the appearing additional hysteresis loops.
At the same time, such systems produce non-trivial magnetization patters, which are quite hard to study theoretically in some cases. 
Machine learning magnetization patterns look like a promising approach.

Here we use the generalized version of the `learning by confusion' scheme~\cite{Nieuwenburg2017} for analyzing magnetic patterns in the ferrimagnetic GdFeCo alloy.
By applying the 'learning by confusion' scheme, we obtain 2D $W$-a shaped surface that characterizes a triple phase transition point of the GdFeCo alloy.

\section{Magnetization patters in GdFeCo films}

We start our consideration from the analysis of the thermodynamic potential of the system, which has the following form: 
\begin{multline}
	\Phi =
	-x\bm{M}_f\bm{H}_{\rm eff}
	-(1-x)\bm{M}_d\bm{H}\\
	- xK_f\left(\frac{\bm{M}_f}{M_f}\bm{Z}\right)^2
	-(1-x)K_d\left(\frac{\bm{M}_d}{M_d}\bm{Z}\right)^2,
\end{multline}
where
$x\in [0, 1]$ is the stechiometric coefficient in the Gd\textsubscript{x}(FeCo)\textsubscript{1-x}, 
$\bm{H}$ is the external field,
$\bm{M}_{f,d}$ are the magnetizations of $4f$-(Gd) and $3d$-(FeCo) sublattices, correspondingly,    
$\bm{H}_{\rm eff}=\bm{H} - \lambda\bm{M}_d$ is the effective magnetic field from the external field and the FeCo sublattice, and 
$\bm{Z}$ is an anisotropy vector.
This thermodynamic potential can be derived from the spin Hamiltonian of a two-sublattice (RE-TM) ferrimagnet~\cite{Davydova2019,Zvezdin1985}.

GdFeCo films have a ferrimagnetic ordering with two sublattices: $4f$-(Gd) and $3d$-(FeCo).
Since the exchange interaction between rare-earth ions are relatively weak, it can be ignored, while the $d$-sublattice has ferromagnetic ordering.
As a result, $4f$ sublattice behaves like a paramagnet in the effective field $\bm{H}_{\rm eff}$, which is created by the $f-d$ exchange interaction of antiferromagnetic type and the external field.
Thus, at relatively low external fields magnetizations of sublattices are oriented antiparallel. However, at high fields the magnetic structure undergoes a spin-flop transition and instead of antiparallel orientation of sublattices the canted phase takes place.

Then let us introduce $\theta_d$ as an angle between $\bm{M}_d$ and $Z$.
GdFeCo films demonstrate perpendicular magnetic anisotropy, which is common for both sublattices~\cite{Ding2013, Leamy1979}. 
We set external magnetic field H to be perpendicular to the film plane ($\bm{H} \parallel \bm{Z}$). Thus, we obtain:
\begin{multline}
	\Phi =
	-x\bm{M_f} \bm{H}_{\rm eff}-(1-x)M_d H\cos\theta_d\\
	- xK_f\left(\frac{\bm{M}_f}{M_f}\bm{Z}\right)^2
	-(1-x)K_d\cos^2\theta_d.
\end{multline}
We assume that the $f$-subblatice is saturated and the orientation of $\bm{M_f}$ is completely defined by the effective field $\bm{H}_{\rm eff}$ ($M_f \parallel \bm{H}_{\rm eff}$).
Then we can set $\bm{M}_f = \chi_f \bm{H}_{\rm eff}$ and arrive at the following expression:
\begin{equation}\label{eq:thermodynamic potential}
	\frac{\bm{M}_f}{M_f}\bm{Z} =
	\frac{\bm{H}_{\rm eff}}{H_{\rm eff}}\bm{Z} =
	\frac{\bm{H} - \lambda\bm{M}_d}{H_{\rm eff}}\bm{Z}=
	\frac{H - \lambda M_d \cos\theta_d}{H_{\rm eff}}.
\end{equation}

\begin{figure}
	\centering
	\includegraphics[width=0.5\textwidth]{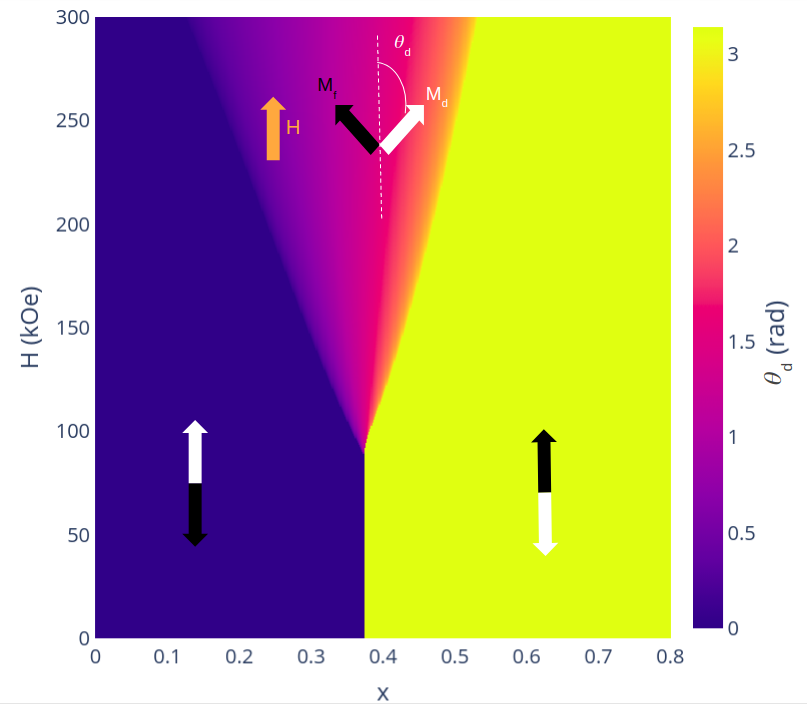}
	\vskip -4mm
	\caption{Phase diagram of the GdFeCo film, which is obtained by the numerical minimization. The colour depicts the value of $\theta_d$ for each point $(x, H)$. Blue and yellow regions represent collinear phases. The transitional region between two collinear phases is corresponding to the canted phase. Black and white arrows show orientation of $f$- and $d$- sublattices in relation to the magnetic field direction.}
	\label{fig:phase}
\end{figure}

Defining the last expression in Eq.~(\ref{eq:thermodynamic potential}) as $\cos\theta_f$ we obtain the following final form of the thermodynamic potential~\cite{Davydova2019}:
\begin{multline}
	\Phi = -xM_fH_{\rm eff} -
	(1-x)M_dH\cos\theta_d -\\
	xK_f\cos^2\theta_f -
	(1-x)K_d\cos^2\theta_d.
\end{multline}

The phase of the system is controlled two parameters: $x$ and $H$, whereas the energy of the system can be obtained by minimizing $\Phi$ with respect to the angle $\theta_d$.
The phase transition can be identified by the observation of the change of a leap of angle $\theta_d$.
Therefore, the phase diagram of the system in principle can be obtained by the corresponding numerical procedure. 
Because at microscopic level magnetic properties of amorphous alloys are spatially non-uniform here and below we restrict ourselves: 
$\mu_{B} = 9.27\cdot10^{-21} \mathrm{erg}\cdot \mathrm{G}^{-1}$,
$M_{f}=10 \mu_{B}$,
$M_{d}=5 \mu_{B}$,
$K_{d,f} = 2.55 \cdot 10^{-16}$ erg,
$\lambda = 10^5 \cdot \mu_{B}^{-1}$.
We illustrate in Fig.~\ref{fig:phase} the phase diagram of the system, which is obtained by the numerical minimization. 

The numerical procedure works, but it is quite computationally intensive even for the idealized system, which is considered above.
In the case of the presence of additional fluctuations of the anisotropy vector $\bm{Z}$ for different cites or with the increase of the system complexity, the scheme that is presented above would experience difficulties. 
Therefore, it is interesting to apply alternative approaches for finding phase transitions in this class of systems.
Below we present the results of the identification of phase transitions with the use of `learning by confusion' scheme~\cite{Nieuwenburg2017}.

\begin{figure*}[htbp]
\includegraphics[width=1\linewidth]{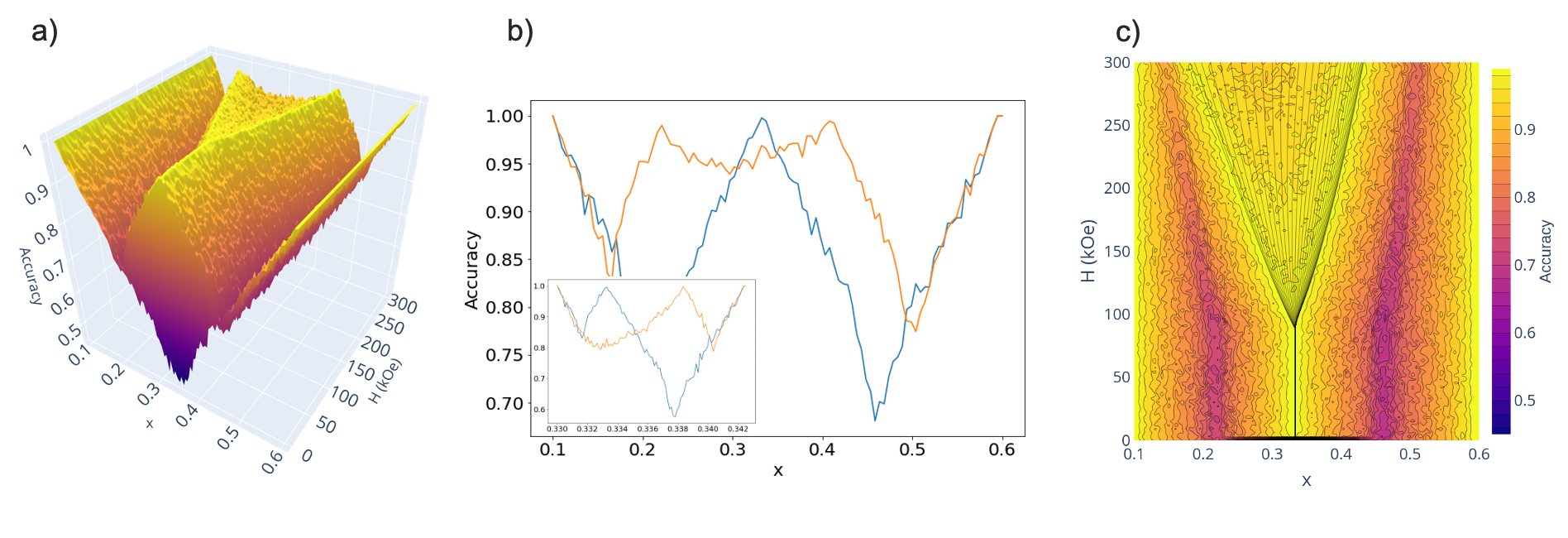}
\vskip -6mm
\caption{Results of the implementation of the `learning by confusion' scheme for the region $x \in [0.1, 0.6]$ and $H \in [0, 3 \times 10^5]$. 
In \textbf{a.} the 3D view of a $W$-shape structure is presented. Lines of phase transition are clearly observed as a ridge on the accuracy surface. 
In \textbf{b.} two slices that represent $W$-shape for particular fixed points of the magnetic field $H$. 
Despite the presence of fluctuations, peaks with high accuracy values are easily distinguished. 
In \textbf{c.} the heat-map of the `learning by confusion' scheme (colored) and numerical calculations of phase transition lines (black).}  
\label{fig:scheme}
\end{figure*}

{\it Learning by confusion}.
For identifying phase transition we use a generalization of the machine learning technique, which is known as `learning by confusion'~\cite{Nieuwenburg2017}.
It is assumed that the phase of the system depends on a critical parameter $c$, which lies in the range $[c_0, c_1]$, so that if $c=c_0$ then the system is in phase `1' and if $c=c_1$ the system is in phase `2'. 
The task is to define the critical value $c^{*}$, which correspond to the transition between two distinct phases. 
This task is difficult especially for complex many-body quantum systems without any prior information about the proper phase attribution.
The `learning by confusion' scheme suggests the following approach. 
We can choose a parameter $c'\in[c_0, c_1]$, and set label `1' to all the data with parameter value less then $c'$ and label `2' to the rest.
In other words, we try to confuse our algorithm by proposing random values of the critical point. 
If our guess of the critical point is correct, the algorithm would learn the hidden properties and predict phase attribution with the highest accuracy.
Otherwise, the algorithm would be confused, would not learn any patterns and fail during the prediction phase.
Applying this procedure on the grid of $c'\in[c_0, c_1]$ we would observe the $W$-shape where the central peak would depict the critical point.
If the system has more than two transition points so that the number of peaks increases, respectively.

Here we form input data for the scheme as follows.
We choose a value of one of the parameters, let say, $x^{*}$ and find the corresponding value of $\theta^{*}$ that provides the minimum of the energy.
In order to safe computational resources, instead of solving the minimization problem of the grid, we sample $n_{\theta_d}$ angles $\theta_d \sim \mathcal{U}(0, \pi)$ for each set of parameters ($H$, $x$).
Then we calculate energy values in each dot and find the angle $\theta^{*}$, which minimizes the energy.

As the next stage, we start the implementation of the `learning by confusion' scheme. 
First, we split all the data that we obtained into two parts, which we label as \texttt{train set} and \texttt{test set}.
The we confuse the machine learning algorithm by proposing some critical values of $x$, train it on the basis of the \texttt{train set}, and evaluate the performance using the \texttt{test set}. 
We expect to obtain the true transition point at the value $x_{\rm crit}$ for whose the machine learning algorithm achieve the best performance. 
In the original approach, neural networks have been used in order to perform the confusion scheme. 
Here as an alternative approach we use the \texttt{XGBoost} algorithm from the \texttt{sklearn} library, which is a powerful technique on the basis of decision trees.
As the result of the scheme, we obtain the 2D $W$-shaped surface that allows identifying clearly phase transitions between various magnetic phases, see Fig.~\ref{fig:scheme}a.
We also demonstrate 1D $W$-shape for the case of several fixed values of the magnetic field [Fig.~\ref{fig:scheme}b], where one can see the perfect accuracy of the machine learning scheme.
The region of interest for a narrow set of the parameters is shown in Fig.~\ref{fig:scheme}c.

We note that the results are in the perfect agreement with the already obtained numerical results (see Fig.~\ref{fig:phase} ), however the latter mentioned are more computationally intensive. 
In order to eliminate numerical effect that possibly appear we launch the algorithm several times and use averaged values. 
However, the obtained error bars are not significant and do not change the physical picture. 

{\it Conclusion}.
In this work, we have demonstrated the possibility to use machine learning methods for analyzing magnetic patterns in ferrimagnets and identifying phase transitions. 
In particular, we have used the `learning by confusion' scheme for obtaining phase transitions with rather high precision. 
We have demonstrated that the obtained results are in the perfect agreement with the already obtained numerical results, however the latter mentioned are more computationally intensive. 

The next step of the research is to apply this method for more and more complex magnetic systems, which cannot be easily described by analytical models and require intensive numerics in order to find the phase diagram. 
We expect that these research are of interest for potential applications.

\medskip

{\it Acknowledgments}.
This work is supported by the Russian Science Foundation (20-42-05002).

\end{document}